\documentclass[11pt]{article}
\def\D{\Delta}
\def\d{\delta}
\def\L{\Lambda}
\def\l{\lambda}

\def\g{\gamma}
\def\e{\epsilon}

\def\o{\omega}
\def\i{\iota}
\def\a{\alpha}
\def\b{\beta}
\def\cd{\cal D}
\def\cj{\cal J}
\def\cl{\cal L}
\def\m{\mu}
\def\th{\theta}

\def\ha{\frac12}
\def\dim{\textrm{dim}}
\def\det{\textrm{det}}
\newcommand{\be}{\begin{equation}}
\newcommand{\ee}{\end{equation}}
\newcommand{\bea}{\begin{eqnarray}}
\newcommand{\eea}{\end{eqnarray}}

\begin{document}

\bigskip
\bigskip
\begin{center}
\bf{\Large Spin Foam Models from the Tetrad Integration}\footnote{Based on the talk presented at the ERE05 meeting, September 6-10, 2005, Oviedo}
\end{center}

\bigskip
\bigskip
\begin{center}
A. MIKOVI\'C\\
Departamento de Matem\'atica,
Universidade Lus\'ofona de\\ Humanidades e Tecnologias,
Av. do Campo Grande, 376\\ 1749-024 Lisbon, Portugal\\

\bigskip
E-mail address: amikovic@ulusofona.pt
\end{center}

\bigskip
\bigskip
\centerline{{\bf Abstract}}

{\small We describe a class of spin foam models of four-dimensional quantum gravity which is based on the integration of the tetrad one-forms in the path integral for the Palatini action of General Relativity. In the Euclidian gravity case this class of models can be understood as a modification of the Barrett-Crane spin foam model. Fermionic matter can be coupled by using the path integral with sources for the tetrads and the spin connection, and the corresponding state sum is based on a spin foam where both the edges and the faces are colored independently with the irreducible representations of the spacetime rotations group. }


\bigskip
\bigskip
\bigskip

The approach of defining a quantum theory of gravity by using a path integral quantization has been revitalized by the appearance of the idea of spin foams \cite{b}. A spin foam model can be described as a lattice gauge theory for a BF theory, and although a BF theory is a topological theory, the Palatini action of General Relativity (GR) can be represented as a constrained BF theory, where the two-form $B$ is a wedge product of the spacetime tetrade one-forms. This then leads to the idea that the GR path integral could be defined as a modification of the path integral for a topological theory. This was the approach used for the construction of the Barret-Crane (BC) models \cite{bce,bc}, which culminated when a finite partition function for GR was constructed for any non-degenerate triangulation of the spacetime manifold \cite{cpr}. However, it was soon realized that one can obtain several finite BC models with different convergence properties \cite{baecr}. This ambiguity is a problem because it is still not clear which one of these has GR as the classical limit. The source of the ambiguity is the fact that the edge amplitudes of the dual two-complex cannot be fixed in the BC quantization procedure, which then leads to many possible models. 

Another problem with the BC type models is that it is difficult to couple matter, especially fermions, because matter fields couple to the tetrades and it is often impossible to rewrite the matter actions coupled to gravity as functionals of the $B$ and matter fields only. One can couple matter to BC models algebraically \cite{amm}, but the algebraic constraints are not strong enoguh to determine the exact matter amplitudes.

These problems of the BC approach suggest that one should try to find a spin foam model which is based on the integration of the tetrade fields in the GR path integral. Such an approach should be feasible because the Palatini action is quadratic in the tetrads, so that the path integral over the tetrads is Gaussian. 

Let us consider the Palatini action 
\be S  =\int_M \e_{abcd}\, e^a \wedge e^b \wedge R^{cd} = \int_M \langle e^2  R \rangle \,d^4 x \quad,\ee
where $M$ is the spacetime manifold, $e^a$ are the tetrad one-forms, $R^{ab} = d\o^{ab} + \o^a_c \wedge\o^{cb}$ is the curvature two-form, $\o^{ab}$ is the spin connection one-form and $\e_{abcd}$ is the totally antisymmetric symbol ($\e_{0123}=1$). The corresponding path integral can be rewritten formally as
\be Z=\int {\cal D} \o \, {\cal D} e \,e^{i\int_M \langle e^2 R \rangle \,d^4 x} = \int {\cal D} \o \, (\det\, R )^{-1/2}\quad, \label{fpi}\ee 
where $(\det\, R)^{-1/2}$ denotes the result of the integration of the tetrads.

The formal expression (\ref{fpi}) suggests that one may try to define $Z$ on a triangulation of $M$ as
\be Z= \int \prod_l dA_l \prod_f (\det \, F_f )^{-1/2}= \int \prod_l dg_l \prod_f 
\D ( g_f)\quad,\ee
where $A_l =\int_l \o$, $g_l =e^{A_l}$, $R_f =F_f =\int_f R$, $\det \,F = (\e^{abcd}F_{ab}F_{cd})^2$,
\be g_f = e^{F_f}=\prod_{l\in\partial f}g_l \quad,\quad 
\D (g_f ) =(\det \, F_f )^{-1/2}\quad,\ee 
and the indices $l$ and $f$ stand for the edges and the faces of the dual two-complex of the triangulation. The group function $\D(g)$ should be gauge invariant, so that we take
\be \D (g) = \sum_\L \D(\L )\,\chi_\L (g) \quad,\ee
where $\chi_\L (g)$ is the character for an irreducible representation (irrep) $\L$, and the sum is over all irreps of a given category (finite-dimensional or unitary). This then implies that 
\be \D(\L)= \int_G dg \,\bar\chi_\L (g)\, \D(g)\quad.\label{pwi}\ee 

By using the formula
\be \int_G dg\, D^{(\L_1)\b_1}_{\a_1}(g)\cdots D^{(\L_4)\b_4}_{\a_4}(g) = \sum_\iota C^{\L_1\cdots\L_4 (\iota)}_{\a_1 \cdots \a_4}\left(C^{\L_1\cdots\L_4 (\iota)}_{\b_1 \cdots \b_4}\right)^* \quad, \label{fi}\ee
where $C^{\L_1\cdots\L_4 (\iota)}_{\a_1 \cdots \a_4}$ are the components of the intertwiners $\iota$ for the tensor product of four irreps and $D^{(\L)}(g)$ are the corresponding representation matrices, we will obtain a state sum of the form
\be Z= \sum_{\L_f,\iota_l} \prod_f \D(\L_f) \prod_v A_v
(\L_f,\iota_l) \quad,\label{grss}\ee
where the vertex amplitude $A_v$ is given by the evaluation of the pentagon spin network, which in the $SU(2)$ case is known as the $15j$ symbol. The state sum (\ref{grss}) is of the same form as in the case of the topological theory given by the BF action; however, the weights we put on the faces are not $\dim\,\L_f$ but the functions $\D(\L_f)$.

These new weights are given by the integrals which are generically divergent, due to 
$\det\, F_f  = 0$ configurations, so that some kind of regularization must be used. In order to do this, let us write the Lie algebra element $F_f$ as
\be F_f = \vec E_f \cdot \vec K + \vec B_f \cdot \vec J \quad,\ee
where $\vec K$ are the boost generators, while $\vec J$ are the spatial rotations generators. The $so(4)$ Lie algebra is a direct sum of two $so(3)$ algebras, and a basis of this decomposition is given by 
\be \vec {\cj}_\pm = \frac12 ( \vec J  \pm \xi\vec K ) \quad,\label{lrd}\ee
where $\xi =1$ in the Euclidian case and $\xi = i$ in the Minkowski case. From (\ref{lrd}) it follows that 
\be \det\, F = (\vec E \cdot \vec B )^2 = \frac{\xi^2}{16}((\vec E_+ )^2 - 
(\vec E_- )^2 ) \quad,\ee
where $\vec E_\pm = \vec B \pm \frac{1}{\xi}\vec E$. Note that $\vec E_\pm$ are real in the Euclidian case, while in the Minkowski case are complex conjugates ($\bar E_+ = E_-$). The group function $\D(g_f)$ is then given by the expression
\be \D(g)= \frac{4}{\xi}((\vec E_+ )^2 - (\vec E_- )^2 )^{-1} \quad,\quad g=e^{\vec E_+ \cdot \vec{\cj}_+ + \vec E_- \cdot \vec{\cj}_-} =g_+ \,g_-\quad.\label{gfe}\ee

Since (\ref{gfe}) has a structure of the relativistic momentum square, and the integral (\ref{pwi}) is an essentially a Fourier transform, one can use the $i\e$ regularisation from QFT \cite{amtet}, so that one obtains
\be \D(j,l) = -{1\over 2\xi} [2\th (l-j ) - \th (l -j +1)  -\th(l-j-1 )] \quad,\label{pgd}\ee
where $j$ and $l$ are the $SU(2)$ spins. The formula (\ref{pgd}) implies that the non-zero coeficients are the ones with $l-j=0$ or $l-j=\pm 1$, 
so that one obtains a weight which is concentrated around the simple irreps $(j,j)$. Hence the model can be considered as a generalization of the Barrett-Crane model.

Including matter and the cosmological constant term requires the evaluation of the path integral
\be Z=\int {\cd}e {\cd}\o {\cd}\psi \exp\left( i \int_M \left(\langle e^2 R \rangle +\l\langle e^4 \rangle\right)d^4 x + iS_m [\psi ,e,\o ] \right) \quad,\label{polpi}\ee
where $\l$ is the cosmological constant, $S_m = \int_M d^4 x \,{\cl}_m$ and ${\cl}_m$ is a function of the tetrads, spin connection, matter fields $\psi$ and their derivatives. Note that in the case of spin-half fermions ${\cl}_m$ is  a polynomial in $e$ and $\o$ given by
\be S_m = \int_M \e_{abcd}\,e^a \wedge e^b \wedge e^c \wedge \bar\psi \left(\g^d \left( d + \ha\o_{rs}\g^r\g^s \right)+ me^d \right)\psi \quad,\ee
where $\g^a$ are the Dirac gamma matrices and $m$ is the fermion mass. Hence the path integral (\ref{polpi}) can be evaluated at least perturbetively. 
 
When ${\cl}_m$ is a polynomial of the fields and their derivatives, the path integral (\ref{polpi}) can be evaluated perturbatively via
\be Z= \lim_{j,J,{\chi}\to 0} e^{i\l\int_M \langle (\d/\d j)^4 \rangle \,d^4 x + iS_m [-i\d/\d {\chi} ,-i\d/\d j ,-i\d/\d J ]} Z_0 [j,J,\chi ] \quad,\ee 
where
\be  Z_0 [J,j,{\chi}]=\int {\cal D}e \, {\cal D}\o \,{\cal D}\psi \, e^{ i\int_M  \left(\langle e^2 R \rangle+ J_{ab}\o^{ab} +j_a e^a + {\chi}^\a \psi_\a  \right)\,d^4 x} \label{gfm}\ee
is the generating functional. $Z_0$ is essentially the gravitational path integral with the sources since the matter integration in (\ref{gfm}) gives a delta function $\d(\chi)$. The tetrade path integral is Gaussian, so that we need to define
\be Z_0 [J,j] = \int {\cal D} \o \, e^{i\int_M d^4 x \,J_{ab}\o^{ab}} (\det R )^{-1/2} e^{-i\int_M d^4 x\int_M  d^4 y 
\langle j(x) R^{-1}(x,y)j(y) \rangle /4}\quad. \label{pifs}\ee

This expression can be defined on a triangulation of $M$ along the lines of the $J=j=0$ case. However, when the sources are present a more intricate state sum will appear. Guided by the expression (\ref{pifs}) we will define $Z_0$ as
\be Z_0 (J,j)= \int \prod_l dg_l \,\m(g_l ,J_l )\prod_f \D(g_f ,j_\e ) \quad,\label{gis}\ee
where
\be \m(g_l ,J_l ) = e^{iTr(\o_l J_l )} \quad,\quad \D(g_f ,j_\e )= \D(g_f) e^{-i \langle j_\e F_f^{-1} j_{\tilde\e}\rangle/4 } \quad,\ee
and the subscripts $\e$ and $\tilde\e$ denote two edges of the triangle dual to the face $f$. 

The functions $\m$ and $\D$ can be expanded as
\be\m (g_l,J_l) = \sum_{\L_l} \m (\L_l ,J_l) D^{(\L_l)}(g_l) \quad,\quad 
 \D(g_f ,j_\e ) = \sum_{\L_f} \D (\L_f ,j_\e) \chi_{\L_f }(g_f)\quad.\ee

The group integrations in (\ref{gis}) can be performed by using the analog of the formula (\ref{fi}) for the tensor product of five irreps. One then obtains a state sum
\be Z_0(J,j)=\sum_{\L_f , \L_l  ,\i_l} \prod_f \D (\L_f , j_\e)  
\prod_l \m (\L_l ,J_l) \prod_v A_v (\L_f , \L_l, \i_l ) \quad.\ee
This is a novel spin foam state sum, because it involves a dual 2-complex whose edges and faces are independently colored with the irreps of the group. $Z_0(J,j)$ can be understood as an amplitude for a Faynman diagram given by a five-valent graph whose edges carry the irreps $\L_l$ and the loops carry the irreps $\L_f$. The edges have propagators $\m (\L_l ,J_l)$ and each loop carries a weight $\D (\L_f, j_\e)$, while the vertex amplitudes $A_v$ are given by the evaluation of the pentagon spin network with five external edges, where the internal edges carry $\L_f$ irreps, while the external edges carry $\L_l$ irreps.

In order to have a physical model, the simplex weights should be defined for the Minkowski case. One way to do this would be to perform an analytic continuation of the Euclidian weights (\ref{pgd}) such that $\xi \to i$. In this approach one would work with the same category of representations as in the Euclidian case, i.e. finite-dimensional $SU(2,{\bf C})\times SU(2,{\bf C})$ representations, so that the Minkowski weights will be the Euclidian weights times the appropriate factors of $i$. An alternative approach would be to use the category of unitary $SL(2,{\bf C})$ representations. In this case the irreps are infinite-dimensional and can be labeled as $(j,\rho)$ where $2j\in\bf {\bf Z}_+$ and $\rho\in{\bf R}_+$. 

An important next step is to study the convergence of the state sum in the Euclidian and the Minkowski case. Note that if the Euclidian state sum turns out to be divergent, it can be regularized by passing to the quantum group at a root of unity, which is usually done in the case of topological spin foam models. However, since our model is non-topological, using a quantum group regularization is not necessary, so that one can use alternative regularizations, for example a gauge fixing procedure for spin foams \cite{fr}.

As far as the semiclassical limit is concerned, this is still an unsolved problem for quantum gravity spin foam models. The difficulty is that in the case of non-topological models the partition function state sum is triangulation dependent. This is an obstruction for finding the smooth-manifold limit. One then needs to study triangulations with increasing number of simplexes. Hopefully one could then extract an effective diffeomorphism invariant action. However, a technique must be developed in order to do this. 

Another problem is that the formula
\be (\det\, R)^{-1/2} = \prod_f (\det\, R_f )^{-1/2} \ee
is an approximation. By replacing $R$ with $R^*$ in the Palatini action one obtains a topological gravity theory, while the corresponding state sum has the same weights as in the non-topological case. One can better understand the model by analyzing the $(\det\, R)^{-1/2}$ operator and the corresponding state sum for the simplest regular triangulation of the four-sphere by six four-simplices\footnote{The corresponding dual one-complex is the five-valent hexagon graph consisting of six verticies and 15 edges.}.

\end{document}